# The controversial early brightening in the first half of 20th century: a contribution from pyrheliometer measurements in Madrid (Spain)

M. Antón, J.M. Vaquero and A.J.P. Aparicio

Departamento de Física, Universidad de Extremadura, Badajoz, Spain.

Corresponding author: Manuel Antón, Departamento de Física, Universidad de Extremadura, Badajoz, Spain.

**Phone**: +34 924 289536

**Fax:** +34 924 289651

**E-mail**: mananton@unex.es

**ABSTRACT**

A long-term decrease in downward surface solar radiation from the 1950s to the 1980s ("global dimming") followed by a multi-decadal increase up to the present ("brightening") have been detected in many regions worldwide. In addition, some researchers have suggested the existence of an "early brightening" period in the first half of 20th century. However, this latter phenomenon is an open issue due to the opposite results found in literature and the scarcity of solar radiation data during this period. This paper contributes to this relevant discussion analyzing, for the first time in Southern Europe, the atmospheric column transparency derived from pyrheliometer measurements in Madrid (Spain) for the period 1911-1928. This time series is one of the three longest dataset during the first quarter of the 20th century in Europe. The results showed the great effects of the Katmai eruption (June 1912, Alaska) on transparency values during 1912-1913 with maximum relative anomalies around 8%. Outside the period affected by this volcano, the atmospheric transparency exhibited a stable behavior with a slight negative trend without any statistical significance on an annual and seasonal basis. Overall, there is no evidence of a possible early brightening period in direct solar radiation in Madrid. This phenomenon is currently an open issue and further research is needed using the few sites with available experimental records during the first half of the 20th century.

## 1. INTRODUCTION

The incoming solar radiation is the main factor controlling the energy budget of the earth-atmosphere system (e.g. Stephens et al., 2012; Wild et al., 2013). Hence, studies about long-term records of solar radiation are required for several topics such as climate models, agriculture, water resources and solar energy applications.

Numerous papers have reported a widespread decrease in surface solar radiation from the 1950s to the 1980s, a phenomenon called global dimming (e.g. Stanhill and Cohen, 2001 and references therein; Liepert 2002; Ohmura, 2009). In addition, a partial recovery of surface radiation values has been detected at many sites from middle 1980s to present, being known as brightening period (e.g. Hatzianastassiou et al., 2005; Wild et al., 2005; Sanchez-Lorenzo et al., 2013). Unfortunately, measurements of solar radiation for the period before the 1950s only exist for a limited number of locations (e.g. Gilgen et al., 1998; Ohmura, 2006; 2009). From these scarce datasets, some authors have suggested an "early brightening" in the first part of the 20$^{th}$ century (Wild, 2009). For example, Ohmura (2006, 2009) reported that global (direct+difuse) solar radiation generally increased in Western Europe from the 1920s to the 1950s, although only five stations have records available for this period. Equally, Lachat and Wehrli (2012, 2013) analyzed pyrheliometer measurements at Davos (Switzerland) from 1909 to 2010. They identified a period of early brightening up to 1929 with a slight positive trend in the atmospheric transmission (Lachat and Wehrli, 2012), although this increase is not related to changes in aerosol transmission (Lachat and Wehrli, 2013). Stanhill and Cohen (2005, 2008) studied century-scale changes in solar forcing at the Earth's surface in the U.S. and Japan since the end of the 19$^{th}$ century from measurements of sunshine duration, with a significant positive linear trend from about the 1900s to the 1940s.

However, other authors found no significant increase of solar radiation in the first half of the 20th century. For instance, Hoyt (1979a) reported no significant long-term trends in atmospheric transmission between 1923 and 1957 using pyrheliometer records at four locations over the world. Roosen and Angione (1984) showed that the intensity of the direct solar beam was especially steady from the mid-1930s to the late-1940s at Mount Montezuma (Chile) and Table Mountain (California). In a more recent study, Ohvril et al. (2009) analyzed multi-decadal variations in atmospheric column transparency based on measurements of direct solar radiation at several locations in the North and East of Europe for the period 1906-2007. These authors suggested that global dimming began as early as 1945, but without any significant variation in the atmospheric transparency between 1910s and 1940s. Finally, Sanchez-Lorenzo and Wild (2012) reconstructed all-sky global radiation from 1885 to 2010 in Switzerland using a homogenous dataset of sunshine duration series. They found no evidence of a possible early brightening from late 19th century to the 1930s, although a brief and strong increase was observed in the 1940s in line with other areas of Western Europe (Sanchez-Lorenzo et al. 2008).

Therefore, due to the limited number of sites with records extending back into the first half of the 20th century, the existence of an "early brightening" period is an open issue. In this framework, the present paper contributes to this relevant discussion analyzing, for the first time, the evolution of the atmospheric column transparency derived from pyrheliometer measurements of broadband direct irradiance in Madrid (Spain) for the period 1911-1928.

## 2. SITE, INSTRUMENT AND DATA

The Astronomical Observatory of Madrid (AOM) was founded in 1785. However, the ongoing systematic astronomical observations did not begin until the second half of the nineteenth century due to social and political instabilities in the country. In 1876, AOM staff began the solar observing program carried out during those early years drawing and counting sunspots for the computation of the Sunspot Number as occurred in many other observatories around the world (Vaquero, 2007). This solar program was expanded to other types of observations such as solar protuberances, solar flocculi, and pyrheliometer measurements.

Pyrheliometer measurements began in 1903 and were abandoned in 1934. The AOM had three Ångström electrical compensation pyrheliometers made by Rosse (Uppsala) with reference numbers 25, 136 and 782. The observations were started with the instrument number 25, which was replaced by the instrument number 136 in September 1910. The instrument number 136 was used until February 1929 when it was replaced due to failure by the instrument number 25. The instrument number 782 was used only for observation campaigns outside Madrid, particularly in summer months. Thus, AOM staff made pyrheliometer observations from the Observatory building, placed in the centre of Madrid in the South part of the Retiro Park, in clear sky conditions for over thirty years. The results of these observations were published in the "Anuario del Observatory de Madrid" and the "Boletín Astronómico del Observatorio de Madrid" (López Arroyo, 2004).

In this work, we have collected all published data. However, we only use only data by the pyrheliometer number 136 (September 1910-February 1929), since the data collected using the instrument number 25 are highly variable due to a strong dependence of the instrumental constant with temperature. To our knowledge, this series can be considered as one of the three longest continuous pyrheliometer dataset during the first quarter of 20th century in Europe together with Davos, Switzerland (Lachat and Wehrli, 2012, 2013) and Pavlovsk, Russia (Ohvril et al., 2009) and the only one in Southern Europe.

## 3. ATMOSPHERIC COLUMN TRANSPARENCY

In this work, the atmospheric column transparency is characterized by the Atmospheric Integral Transparency Coefficient (AITC) defined as

$$AITC_m = \left(\frac{I_m}{I_0}\right)^{1/m} \quad (1)$$

where $I_m$ is the experimental broadband direct irradiance in cloud-free solar disc at relative optical air mass m, and $I_0$ is the extraterrestrial broadband solar irradiance (1367 W/m$^2$) corrected to the actual Sun-Earth distance. This transparency parameter, also called Bourger atmospheric transmittance coefficient, is derived from the pyrheliometric formula which is based on the Bouguer-Lambert law (e.g. Kondratyev 1969; Kasten 1980). AITC has been used by numerous authors to characterize the atmospheric column transparency in different locations worldwide (e.g. Russak, 1990, 2009; Abakumova et al., 1996; Alados-Arboledas et al., 1997; Olmo et al., 1999; Ohvril et al., 1999, 2009; Kannel et al.,, 2012).

AITC represents the atmospheric column transmission averaged over the entire solar spectrum of direct solar beam which changes throughout the day even in the case of stationary and azimuthally homogeneous atmosphere. Therefore, AITC depends on solar elevation causing the diurnal variation of the atmospheric (broadband) column transmission, phenomenon known as the Forbes effect (Ohvril et al., 1999). This effect is related to the high wavelength-dependence of solar radiation in its way through the atmosphere during the day. Thus, longer wavelengths contribute proportionately more to direct irradiance as the solar elevation decreases due to the substantial attenuation of shorter wavelengths by Rayleigh scattering. Consequently, the peak contributions shift to longer wavelengths where the solar rays become more penetrating and, therefore, the broadband coefficient of transparency increases.

To remove the dependence of AITC on solar elevation caused by the Forbes effect, all their daily values were reduced from the actual optical air mass m to optical air mass m = 2 (solar elevation of 30º) by the expression proposed by Myurk and Okhvril (1990) and subsequently used by different authors (e.g. Alados-Arboledas et al., 1997; Olmo et al., 1999; Ohvril et al., 1999, 2009; Russack, 2009; Kannel et al., 2012):

$$AITC_2 = AITC_m \left( \frac{2}{m} \right)^{\frac{\log AITC_m + 0.009}{\log m - 1.848}} \quad (2)$$

**4. RESULTS AND DISCUSSION**

Firstly, the day-to-day changes in the atmospheric column transparency over Madrid during the study period were analyzed. For this goal, the relative variation in $AITC_2$ was calculated as follows:

$$\left( \Delta AITC_2 \right)_i = 100 \times \frac{\left| AITC_2^{i+1} - AITC_2^{i} \right|}{AITC_2^{i}} \quad (3)$$

where $AITC_2^{i}$ and $AITC_2^{i+1}$ are the Atmospheric Integral Transparency Coefficient for two consecutive days i and i + 1. The mean (median) value of $\Delta AITC$ for all cases was 3.7% (2.2%), which indicates that the atmospheric column transparency over Madrid showed on average small day-to-day changes. Nevertheless, punctual large day-to-day variations were found, reaching the 95$^{th}$ and 99$^{th}$ percentile of $\Delta AITC$ a value of 12.3% and 23.2%, respectively. These large day-to-day fluctuations could be associated with desert dust intrusions from Northern Africa and local aerosol sources like smoke plume from occasional forest fires nearby.

To minimize the effect of day-to-day turbidity fluctuations, monthly mean time-series dataset, M(t), was derived from the averages of the daily $AITC_2$ data. The evolution of these monthly $AITC_2$ values for the period from January 1911 to December 1928 in Madrid is shown in Figure 1 (top). A buildup of $AITC_2$ values during winter can be seen, while a decline is observed through summer. On average (± one standard deviation), maximum monthly $AITC_2$ is reached in December (0.76±0.05), and minimum value in July and August (0.71±0.04), while the annual mean using all the available monthly data has a value of 0.74±0.04. This annual cycle is anticorrelated with cycles of integrated water vapor content and aerosol concentration which are the main responsible of atmospheric transparency variations in the case of an unscreened solar disc (e.g. Hoyt, 1979b; Lachat and Wehrli, 2012, 2013). For instance, Bennouna et al. (2013) analyzed the seasonal evolution of the precipitable water vapor (PWV) at ten Spanish locations from 10-year data collected by ground-based GPS receivers. These authors showed that the PWV presents a clear annual cycle, with a minimum in winter, reaching a maximum at the end of the summer and decreasing during the autumn. Additionally, numerous papers have shown high aerosol load during summer at different Spanish sites related to the seasonal patterns of the desert dust particles which arrive at the Iberian Peninsula from Northern Africa (e.g. Lyamani et al., 2005, 2010; Toledano et al., 2007; Guerrero-Rascado et al., 2008; Valenzuela et al., 2012a, 2012b).

The first step in the analysis of the long-term $AITC_2$ trend must be to deseasonalize the monthly series due to its marked seasonal component as shown in the previous paragraph.. In this work, the annual cycle of $AITC_2$ data (seasonal pattern) was estimated from the best fit of monthly $AITC_2$ values by least squares method using the following function

$$S(t) = a + b \cdot \sin(w \cdot t) + c \cdot \cos(w \cdot t) \quad (3)$$

where t is the time in months, $w = 2\pi/12$, a is the central $AITC_2$ value, and the term $b \cdot \sin(w \cdot t) + c \cdot \cos(w \cdot t)$ represents the seasonal component of the $AITC_2$ variability. The amplitude of this seasonal component was calculated as $\sqrt{b^2 + c^2}$, resulting a value of 0.033±0.003. Finally, the seasonal component of natural atmospheric transparency variations is subtracted from the monthly $AITC_2$ dataset to obtain the deseasonalized monthly time series as

$$D(t) = M(t) - S(t) \quad (4)$$

Figure 1 (bottom) shows the evolution of the deseasonalized monthly time series, D(t), derived from the pyrheliometer dataset in Madrid for the period 1911-1928. In addition, this time series was smoothed with a 12-month running averages in order to make the long-term atmospheric transparency variability more prominent. A marked drop in the atmospheric transparency is appreciated during 1912-1913 related to pollution from the Katmai eruption on 6 June 1912 in central Alaska, considered as the $20^{th}$ century's most powerful volcano and comparable to the famous Krakatau (1883) (see table 1 in Sato et al., 1993). It should be noted that the influence of this volcano on $AITC_2$ values at Madrid was detected up to 12 months after the eruption, affecting mainly the second half of 1912 and the first half of 1913 (conditions close to that prevailing in the pre-eruption period were reached in June 1913). Thus, the mean $AITC_2$ anomaly between June 1912 and May 1913 was -0.05±0.02, reaching an extreme value of -0.083 (relative anomaly of 8.3%) in August 1912. In addition, throughout the whole study period, the lowest annual mean $AITC_2$ were reached in 1912 and 1913 with a value of 0.72 for both years. Ohvril et al. (2009) reported annual mean $AITC_2$ values of 0.63 and 0.74 in Pavlovsk (Russia) during 1912 and 1913, respectively. The comparison of $AITC_2$ values between Madrid and Pavlovsk suggests that the Katmai's effects on atmospheric transparency were more pronounced in Northern than Southern sites in Europe during 1912 but with a similar effect during 1913. On the other hand, shorter time series of $AITC_2$ values for several Spanish locations also confirmed a marked degradation in atmospheric column transparency after the El Chichón (1982) and Mount Pinatubo (1991) eruptions (Alados-Arboledas et al., 1997; Olmo et al., 1999).

Outside the period affected by the Kutmai eruption, Figure 1 (bottom) shows a fairly stable behavior without clear visual positive or negative long-term trends. To verify this issue, the linear $AITC_2$ trend is obtained from the linear regression analysis applied on the annual time series (derived from the annual average of the deseasonalized monthly values) for the period 1914-1928 after the Katmai eruption. The slope of the trend line (± standard error) showed a small negative value of $(-7.3 \pm 4.4) \cdot 10^{-3}$ per decade without statistical significance at the 95% confidence level, indicating a slight decrease of the atmospheric transparency throughout the study period. Therefore, there is no evidence of a possible early brightening in atmospheric transparency in Madrid for the period 1914-1928. Due to the limited number of experimental direct solar data during the first quarter of the 20$^{th}$ century, our trend results only can be compared with a few papers. Among them, Lachat and Wehrli (2012) calculated the atmospheric transmission from pyrheliometer data at Davos (Switzerland) from 1909 to 2010. These authors performed a linear trend analysis for the period 1909-1929, giving a small positive slope of $(4.5 \pm 2.9) \cdot 10^{-3}$ per decade, although this result is not statistically significant. Ohvril et al. (2009) showed annual mean $AITC_2$ values at Pavlovsk (Russia) for the period 1906-1936 (see their Figure 1), but not linear trends were reported. Nevertheless, these authors indicated that any significant change in the atmospheric transparency was observed for that period. Similar results have been reported for pyrheliometer data recorded during the second quarter of the 20$^{th}$ century. For example, Hoyt (1979a) showed no long-term trends in atmospheric transmission between 1923 and 1957 using the pyrheliometer measurements of the Smithsonian Astrophysical Observatory. Overall, although it seems that there are some indications for an early brightening in global (direct+diffuse) solar data between 1930s and 1940s (e.g. Ohmura, 2006, 2009; Wild, 2009), there are no clear evidences of a significant brightening in direct solar irradiance during the first half of 20$^{th}$ century.

Finally, in order to detect seasonal differences of the atmospheric transparency trends, four seasonal time series were inferred from the average of the deseasonalized values for each season of the year: winter (December, January, and February), spring (March, April, and May), summer (June, July, and August), and autumn (September, October, and November). Thus, the long-term $AITC_2$ trends were also calculated separately for each season by means of a linear regression analysis applied on each time series. From this seasonal analysis, we also found no statistically significant linear trends with values of $(-8.8\pm7.4)\cdot10^{-3}$, $(-6.6\pm6.1)\cdot10^{-3}$, $(-12.4\pm9.5)\cdot10^{-3}$, and $(-0.3\pm8.9)\cdot10^{-3}$ per decade for winter, spring, summer and autumn, respectively.

## 5. CONCLUSIONS

From the results obtained in this study, we can state that there is no evidence of an early brightening period in atmospheric column transparency from pyrheliometer data in Madrid during the period 1911-1928. Even a small negative trend of this variable (~ $4.5\cdot10^{-3}$ per decade), but not statically significant, has been found for this unique dataset, which it is one of the three longest during the first quarter of $20^{th}$ century in Europe. The long-term evolution of the atmospheric column transparency showed a fairly stable behavior only perturbed by the Katmai volcanic eruption (June 1912) whose effects were substantially detected up to middle of 1913. Nevertheless, the mean annual value of the atmospheric transparency obtained in 1912 (~ 0.72) is significantly higher than the annual value reported in a Russian location for the same year. This result could suggest that the influence of the Katmai volcano was stronger in Northern than Southern regions in Europe.

To sum up, although there are some indications for an early brightening in global solar radiation during the first half of $20^{th}$ century, it is not evident for direct solar radiation. Therefore, the early brightening phenomenon needs further investigation and, due to the lack of experimental solar radiation data before 1950s, the analysis of related variables (e.g, sunshine duration) could be very promising in order to know the decadal variations of global and direct solar radiation.

*Acknowledgments*-- Authors thank the useful comments by M. C. Gallego and A. Sanchez-Lorenzo. Manuel Antón thanks Ministerio de Ciencia e Innovación and Fondo Social Europeo for the award of a postdoctoral grant (Ramón y Cajal). Support from the Junta de Extremadura (Research Group Grant No. GR10131) and Ministerio de Economía y Competitividad of the Spanish Government (CGL2011-29921-C02-01, AYA2011-25945) is gratefully acknowledged.

## FIGURES CAPTIONS

**Figure 1. Top:** Time series of the monthly $AITC_2$ (Atmospheric Integral Transparency Coefficient at m=2) for the period from January 1911 to December 1928 in Madrid**. Bottom:** Time series of the deseasonalized monthly $AITC_2$ (equation 4) together with a 12-month running averages for the period from January 1911 to December 1928 in Madrid.

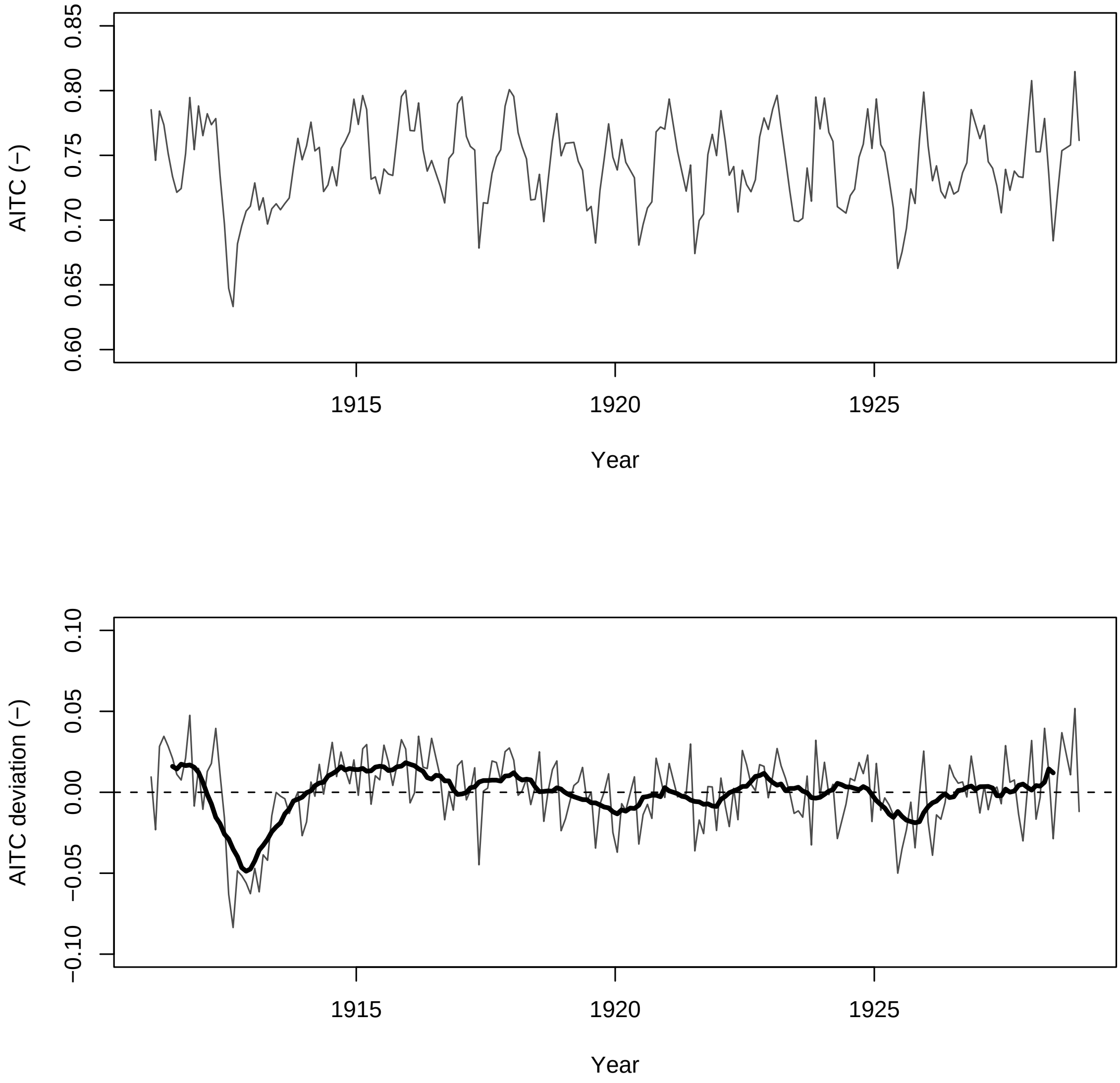

**Figure 1:** Top: Time series of the monthly AITC2 (Atmospheric Integral Transparency Coefficient at m=2) for the period from January 1911 to December 1928 in Madrid. Bottom: Time series of the deseasonalized monthly AITC2 (equation 4) together with a 12-month running averages for the period from January 1911 to December 1928 in Madrid.